\documentclass[runningheads]{llncs}

\usepackage{multicol}
\usepackage{color}

\usepackage{amsmath}
\usepackage{amsfonts}
\usepackage{inputenc}
\usepackage[english]{babel}
\usepackage[]{graphicx}
\usepackage{alltt}
\usepackage[caption=false,font=footnotesize]{subfig}

\usepackage{lineno}

\makeatletter
\RequirePackage[bookmarks,unicode,colorlinks=true]{hyperref}%
   \def\@citecolor{blue}%
   \def\@urlcolor{blue}%
   \def\@linkcolor{blue}%

\def\orcidID#1{\smash{\href{http://orcid.org/#1}{\protect\raisebox{-1.25pt}{\protect\includegraphics{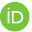}}}}}
\makeatother

\newcommand\blfootnote[1]{%
  \begingroup
  \renewcommand\thefootnote{}\footnote{#1}%
  \addtocounter{footnote}{-1}%
  \endgroup
}

\begin{document}

\title{Automated and Sound Synthesis\\ of 
Lyapunov Functions with SMT Solvers}

\author{
}
%
%
\institute{
}

\author{Daniele~Ahmed\inst{1,2},
Andrea~Peruffo\inst{1}\orcidID{0000-0002-7767-2935}, and 
Alessandro~Abate\inst{1}\orcidID{0000-0002-5627-9093}
}
\authorrunning{D. Ahmed et al.}

\institute{
Department
of Computer Science, University of Oxford, OX1 3QD Oxford, UK\\
\email{ name.surname@cs.ox.ac.uk }
\and
Amazon Inc, London, UK
}


\maketitle
\begin{abstract}
In this paper we employ SMT solvers to soundly synthesise Lyapunov functions that assert the stability of a given dynamical model.  
The search for a Lyapunov function is framed as the satisfiability of a second-order logical formula, 
asking whether there exists a function satisfying a desired specification (stability) for all possible initial conditions of the model.   
We synthesise Lyapunov functions for linear, non-linear (polynomial), and for parametric models. 
For non-linear models, the algorithm also determines a region of validity for the Lyapunov function.  
We exploit an inductive framework to synthesise Lyapunov functions, starting from parametric templates. 
The inductive framework comprises two elements: 
a \textit{learner} proposes a Lyapunov function, and a \textit{verifier} checks its validity - its lack is expressed via a counterexample 
(a point over the state space), for further use by the learner.  
Whilst the verifier uses the SMT solver Z3, thus ensuring the overall soundness of the procedure, we examine two alternatives for the learner: 
a numerical approach based on the optimisation tool Gurobi, 
and a sound approach based again on Z3. 
The overall technique is evaluated over a broad set of benchmarks,   
which shows that this methodology not only scales to 10-dimensional models within reasonable computational time, 
but also offers a novel soundness proof for the generated Lyapunov functions and their domains of validity.   
\end{abstract}


\section{Introduction}
\label{sec:intro} 

\blfootnote{
Tools and Algorithms for the Construction and Analysis of Systems - 26th International Conference, {TACAS} 2020, Proceedings, Part {I},
LNCS 12078, pp. 97--114, Springer, 2020.
\url{
https://doi.org/10.1007/978-3-030-45190-5_6}
}

Dynamical systems represent a major modelling framework in both theoretical and applied sciences: 
they describe how objects move by means of the laws governing their dynamics in time. 
Often they encompass a system of ordinary differential equations (ODE) with nontrivial solutions. 

This work aims at studying the stability property of general ODEs, 
without knowledge of their analytical solution. 
Stability analysis via Lyapunov functions 
is a known approach to assert such property. 
As such, the problem of constructing relevant Lyapunov functions for stability analysis has drawn much attention in the literature \cite{giesl2015review,K15}. 
A brief introduction to the concepts of Lyapunov stability is presented in Section \ref{sec:synthesis}. 
By and large, existing approaches leverage Linear Algebra or Convex Optimisation solutions, 
and are not fully automated nor numerically sound. 

\smallskip

\subsubsection{Contributions}
We apply an inductive synthesis framework, 
known as Counter-Example Guided Inductive Synthesis (CEGIS) \cite{solar2006combinatorial,david2017program} 
and recently employed in a number of control applications \cite{bcABCCDKPK17,ABCCDKKP20,RS15,RS16}, 
to construct Lyapunov functions for linear, polynomial and parametric ODEs, 
and (for non-linear ODEs) to constructively characterise their domain of validity.  
CEGIS, originally developed for program synthesis based on the satisfiability of second-order logical formulae,   
is employed in this work with template Lyapunov functions and in conjunction with a Satisfiability Modulo Theory (SMT) solver \cite{kroening2016decision}.  
Our results offer a formal guarantee of correctness in combination with a simple algorithmic implementation.

The synthesis of a Lyapunov function $V$ can be written as a second-order logic formula 
$
F: = \exists V \ \forall x : \psi  
$, 
where $x$ represents the state variables and $\psi$ represents requirements that $V$ needs to satisfy in order to be a Lyapunov function.

%
The CEGIS architecture is structured as a loop between two components, a ``learner'' and a ``verifier''.
The learner provides a candidate function $V$ and the verifier checks the validity of $\psi$ over the set of $x$; 
if the function is not valid, the verifier provides a counterexample, namely a point $\bar x$ in the state space where the candidate function does not satisfy $\psi$.  
The learner incorporates the generated counterexample $\bar x$, subsequently computes a new candidate function, and passes it back
to the verifier. 

We exploit SMT solvers to (repeatedly) assert the validity of $\psi$, given $V$, over a domain in the space of $x$. 
Satisfiability Modulo Theory (SMT) is a powerful tool to assert the existence of such a function. 
An SMT problem is a decision problem -- a problem that can be formulated as a yes/no question -- 
for logical formulae within one or more theories, e.g. the theory of arithmetics over real numbers. 
The generation of simple counterexamples $\bar x$ is a key new feature of our technique.   

Furthermore, 
in this work we provide two alternative CEGIS implementations: 
1) a numerical learner and an SMT-based verifier, and 
2) an SMT-based learner and verifier. 
The numerical generation of Lyapunov functions is based on the optimisation tool Gurobi \cite{gurobi}, 
whereas the SMT-based one leverages Z3 \cite{de2008z3}. 


\smallskip

\subsubsection{Related Work}  

The construction of Lyapunov functions is recognisably an important yet hard problem, particularly for non-linear ODE models, 
and it has been the objective of classical studies \cite{KB60,K63,LL61}. 
A know constructive result has been introduced in \cite{Z64}, which additionally provides an estimate of the domain of attraction. 
It has led to further work based on recursive procedures.  
Broadly, these approaches are numerical and based on the solution of optimisation problems. 
For instance, linear programming is exploited in \cite{brayton1979stability} to iteratively search for stable matrices inside a predefined convex set, 
resulting in an approximate Lyapunov function for the given model.   
Alternative approximate methods include \cite{giesl2015review} $\varepsilon$-bounded numerical methods, 
techniques leveraging series expansion of a function, 
the construction of functions from trajectory samples, 
and the framework of linear matrix inequalities. 
The approach in \cite{parrilo2000structured} uses sum-of-squares (SOS) polynomials to synthesise Lyapunov functions, however its scalability remains an issue.  
The work in \cite{papachristodoulou2002construction} uses SOS decomposition to synthesise Lyapunov functions for (non-polynomial) non-linear systems: 
the algorithmic implementation is know as SOSTOOLS \cite{prajna2004sostools,papachristodoulou2018sostools}.
\cite{Getal14} focuses on an analytical result involving a summation over finite time interval, under a stability assumption.  
Recent developments are in \cite{H07} and subsequent work, whereas surveys on this topic are in \cite{giesl2015review,K15}. 

In conclusion, existing constructive approaches either rely on complex candidate functions (whether rational or polynomial), 
on semi-analytical results, 
or alternatively they involve state-space partitions (for which scalability with the state-space dimension is problematic) accompanied by correspondingly complex or large optimisation problems.  
These approximate methods evidently lack either numerical robustness, 
being bound by machine precision, or algorithmic soundness:  
they cannot provide formal certificates of reliability which, in safety-critical applications, can be an evident limit.

In \cite{sankaranarayanan2013lyapunov} Lyapunov functions are soundly found within a parametric framework, 
by constructing a system of linear inequality constraints over unknown coefficients. 
A twofold linear programming relaxation is made: it includes interval evaluation of the polynomial form and ``Handelman representations" for positive polynomials. 
Simulations are used in \cite{kapinski2014simulation} 
to generate constraints for a template Lyapunov function, which are then resolved via LP, resulting in candidate solutions. 
Whilst the authors refer to traces as counterexamples, they do not employ the CEGIS framework, as in this work. 
When no counterexamples are found, 
\cite{kapinski2014simulation} further uses dReal \cite{gao2013dreal} and Mathematica \cite{Mathematica} to verify the obtained candidate Lyapunov functions. 
The sound technique, which is not complete, is tested on low-dimensional models with non-linear dynamics. 


The cognate work in \cite{RS15,RS16,RS18} is the first to employ a CEGIS-based approach to synthesise Lyapunov functions. 
\cite{RS15,RS16} focuses on such synthesis for switching control models - a more general setup that ours. 
\cite{RS15} employs an SMT solver for the learner, 
and towards scalability solves an optimisation problem over LMI constraints for the verifier over a given domain (unlike our approach). 
As such, counterexamples are matrices, not points over the state space, and furthermore the use of LMI solvers does not in principle lead to sound outcomes. 
Along the above line, \cite{RS16} expands this approach towards robust synthesis;  
\cite{RS18} instead employs MPC (Model Predictive Control) techniques within the learner to suggest template functions, 
which are later verified via semi-definite programming relaxations (again, possibly generating counterexamples by solving optimisation problems over a given domain).  
Whilst inspired by this line of work, our contribution provides a simple (with interpretable counterexamples that are points over the state space) yet effective (scalable to at least 10-dimensional models) SAT-based CEGIS implementation, which automates the construction of Lyapunov functions and associated validity domains, which is is sound, and also applicable to parameterised models. 


\medskip

The remainder of the paper is organised as follows. 
In Section \ref{sec:formal-methds} we present the SMT Z3 solver  
and the inductive synthesis (IS) framework. 
The implementation of CEGIS, for both linear and non-linear models, is explained in Section \ref{sec:synthesis}. 
Experiments and case studies are in Section \ref{sec:case-study}. 
Finally, conclusions are drawn in Section \ref{sec:concl}.


\section{Formal Verification -- Concepts and Techniques}
\label{sec:formal-methds}

%
In this work we use Z3, an SMT solver, and the CEGIS architecture, 
to build and to verify Lyapunov functions.

\subsection{Satisfiability Modulo Theory}
\label{subsec:SMT}

A Satisfiability Modulo Theory problem is a decision problem formulated within a theory, e.g. first-order logic with equality \cite{clarke2018handbook}. 
The aim is to check whether a first-order logical formula within such theory, referred to as an SMT instance, is satisfied.  
For example, a formula can be the inequality $3x_0 + x_1 > 0$ evaluated within the theory of linear inequalities. 
An SMT solver is a software that checks the satisfiability of an SMT instance, i.e. whether there exists an instantiation of the formula that evaluates to \texttt{True}.
SMT solvers can be useful for function synthesis, 
namely to mechanically construct a function, given requirements on its output. 

\subsection{The Z3 SMT Solver}
\label{subsec:z3}

Z3 \cite{de2008z3,github_z3} is a powerful SMT solver that integrates SAT solvers, 
theory solvers for equalities and interpreted functions, 
satellite solvers for arithmetic, real, array, and other theories, 
and an abstract machine to handle quantifiers. 
Receiving an input formula, 
Z3 represents it as an abstract syntax tree and processes it with its SAT solver core, until it returns \texttt{SAT} if the formula is satisfiable,  \texttt{UNSAT} otherwise. 

\medskip 

\begin{example}[Operation of Z3] 
Consider the formula $a = b \wedge f(a) = f(b)$ in the theory of equality.  
To verify its satisfiability, Z3 constructs a syntax tree, with nodes for each variable ($a$, $b$) and formulae ($a=b$, $f(a)$, $f(b)$, $f(a) = f(b)$).
Once the tree is built, 
Z3 merges $a$ with $b$ and $f(a)$ with $f(b)$ to represent the equality operation and, in order to verify the correctness of the assertion,
applies the congruence rule 
$ \bigwedge_{{i=0}}^{{n-1}} x_i = y_i \Rightarrow f(x_0, \dots x_{n-1}) = f(y_0, \dots y_{n-1})$ to conclude that $a = b\Rightarrow f(a) = f(b)$.
Finally, nodes $a=b$ and $f(a)=f(b)$ are merged and Z3 returns $\mathtt{SAT}$.
\hfill $\square$
\end{example}

\medskip

Of particular interest for the synthesis of Lyapunov functions, is the ability of Z3 to solve polynomial constraints. 
Z3 stores and exactly manipulates algebraic real numbers that are roots of rational univariate polynomials: 
this is done for an algebraic real $\alpha$, by storing a polynomial $p(x)$ for which
$p(\alpha) = 0$ and two rationals $l, u$ such that $p(x) = 0$ for $x \in (l,u)$
if and only if $x = \alpha$. 
In this work, Z3 has been used through its Python APIs, named Z3Py.
An example of a simple assertion verification follows.

\medskip

\begin{example}[Assertion in Z3] 
Consider the (valid) formula
$ x \geq 0 \Rightarrow 3x+1 > 0$.
The code using Z3Py results in:
\begin{alltt}
x = Real(\textsc{\char13}x\textsc{\char13})
s = Solver()
s.add(Implies(x >= 0, 3 * x + 1 > 0))
print(s.check())
\end{alltt}
which evaluates (as expected) to $\mathtt{SAT}$.
\hfill $\square$
\end{example}

\subsection{Inductive Synthesis - CEGIS}
\label{subsec:is}

An approach to solve second-order logic problems, 
such as those characterising the synthesis of Lyapunov functions, is \textit{inductive synthesis} (IS). 
IS infers general rules (or functions) from specific examples (observations),  
entailing the process of generalisation. 
Within the IS procedure, a synthesiser attempts the construction from a (usually small) subset  of the original specifications. 
It then generalises to the complete specification by identifying patterns in the input data. 

\begin{figure}
\centering
\framebox{
\includegraphics[trim={8cm 21.5cm 8cm 4.5cm}, clip=true, width=0.5\textwidth]{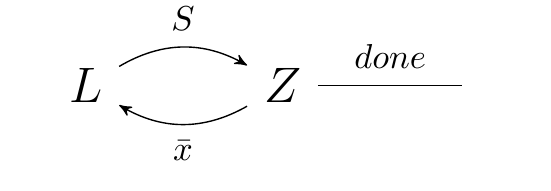}}
\caption{CEGIS-based inductive synthesis. The iterative procedure loops between a learner $L$ and a verifier $Z$. 
$L$ provides a candidate solution $S$ to the verifier $Z$, which asserts its validity or outputs a counterexample $\bar{x}$. 
The learner provides a new solution encompassing also $\bar{x}$. The procedure stops once no counterexamples are found. }
\label{fig:cegis-scheme}
\end{figure}

An exemplar of IS is the CEGIS framework. 
Fig. \ref{fig:cegis-scheme} depicts the relation between its two main components. 
It sets off with a given specification $\psi$ over a set $\mathcal I$ for the synthesis. 
The synthesis engine (a component that will be also denoted as \textit{learner}) provides a candidate solution for $\iota$, a subset of $\mathcal{I}$, the space of possible inputs.
This candidate solution is passed to a second component, called \textit{verifier}, that acts as an oracle: 
either it approves the solution over the entire $\mathcal{I}$, so that the process terminates, or it finds an instance $\bar{x}$ (a counterexample in $\mathcal{I}$) 
where the candidate solution does not comply with the specifications. 
The learner takes $\bar{x}$ and adds it to $\iota$, computing a new (more general) candidate solution for the problem. 
This cycle is repeated. 
Note that this algorithm might not terminate, depending on the structure of $\mathcal{I}$, or might take many cycles to find a proper solution: 
in those instances, tailored candidate solutions and insightful counterexamples are necessary. 
In this work, the IS is implemented using SMT-solvers. 
The verifier finds counterexamples $\bar{x}$ by seeking a witness of the negated formula $\neg\psi$, namely trying to prove that a violation of the formula exists. 
The learner might employ SMT solvers to solve the system of constraints generated by the counterexamples, i.e. to find a valid instance of such constraints, however in general it does not need to be sound, as it is the verifier that guarantees the soundness of the proposed solution. 
Section \ref{subsec:cegis-architecture} illustrates the two CEGIS components, the learner $L$ and the verifier $Z$ in relation to Lyapunov function synthesis. 

\medskip

\begin{example}[CEGIS Operation]
Assume the task is the synthesis of
a function
 $g(x)$ that satisfies the following formula $F(g(x))$: 
$$
\exists \ g(x) \ \forall x \in \mathbb{R} : \psi, \text{ where }
\psi(g(x)) = g(x) + 1 > 0.
$$
The learner $L$ offers an initial (often na\"ive, random or default) candidate, e.g. $g(x) = x$, and passes it to the verifier $Z$. 
The verifier checks the validity of $\psi(x) = x + 1 > 0,$ $\forall x \in \mathbb{R},$  
by searching an instance $\bar{x}$ that might invalidate the formula.  
$Z$ finds that $\bar{x} = -1$ invalidates the formula, thus sends $\bar{x}$ to $L$, 
which incorporates this counterexample to synthesise a new $g(x)$. 
The learner now adds a constraint on the next candidate, as 
$$
C := g(-1) +1 > 0, \ \ \forall x \in \mathbb{R},
$$
such that the new candidate solution satisfies the formula at $\bar{x} = -1$.
The learner now proposes $g(x) = x^2$, which satisfies $C$, and passes it to $Z$.
The verifier searches for a counterexample to $\psi(x^2)$, but cannot find any. 
Thus, it exits the loop with an \texttt{UNSAT} answer, which proves that the synthesised function $g(x) = x^2$ is valid $\forall x \in \mathbb{R}$. 
\hfill $\square$
\end{example}

\section{Automated and Sound Synthesis of Lyapunov Functions via CEGIS and SMT}
\label{sec:synthesis}



Consider a dynamical system $\dot{x} = f(x)$, 
where $f: \mathbb{R}^n \rightarrow \mathbb{R}^n$, 
and assume that the point $x_e \in \mathbb{R}^n$ is an equilibrium, namely such that $f(x_e) = 0$ -- without loss of generality, we assume that $x_e = 0$ (the origin). 
The goal is assessing the stability of such equilibrium point via the synthesis of a Lyapunov function $V(x): \mathbb{R}^n \rightarrow \mathbb{R}$.  
The stability of an equilibrium 
 guarantees that trajectories starting by the equilibrium remain close to it at all times 
(how close can often be quantified, as done later in this work).  
If $V(x)$ fulfils the following two conditions, $\forall x \in \mathcal{D}$, 
\begin{equation}
V(x) > 0,
\quad 
\dot{V}(x) = \nabla V(x) \cdot f(x) \leq 0, 
\label{eq:lyapunov-conditions}
\end{equation}
where $\mathcal{D}$ is a domain of interest containing $x_e$  
then the Lyapunov function ensures boundedness of the trajectories.
In other words, for every initial point in a neighbourhood of $x_e$, the trajectories of the model do not escape from $\mathcal{D}$ 
(with reference to notations introduced above, 
the condition in \eqref{eq:lyapunov-conditions} represents the requirement $\psi$, 
and $\mathcal D$ denotes the set of inputs $\mathcal I$).  
We use the following polynomial expression for the Lyapunov function
\begin{equation}
\label{eq:V=x^i Pi x^i}
V(x) =  \sum_{l=1}^c (x^l)^T \ P_l \ x^l, 
\end{equation} 
where $x^l$ represents the element-wise exponentiation of vector $x$, i.e. element $x(j)$ to the power $l$, $\forall j = 1, \dots, n$; 
$P_l \in \mathbb{R}^{n \times n}$ is a weighting matrix associated with $x^l$, 
and $2c$ is the order of the polynomial function. 
 %
In order to obtain a proper Lyapunov function $V(x)$, 
the synthesiser is asked to verify the specification expressed by the formula
\begin{equation}
\label{eq:synthesis-formula}
F(V(x)): \forall x \in \mathcal{D},  V(x) > 0 \wedge \dot{V}(x) \leq 0.
\end{equation} 
This specification requires the Lyapunov function to be positive definite, and not to increase along the trajectories of the model. 
For linear systems, unless otherwise stated, we consider $\mathcal{D} = \mathbb{R}^n \setminus \{0\}$ and $c=1$, 
as it is known that quadratic functions are sufficient to prove the stability of linear models over the whole state space.  
Formula \eqref{eq:synthesis-formula} keeps the elements of $P$ uninterpreted, and thus they are parameters to be found. 
Notice that the second-order formula 
\begin{equation*}
\exists P \in \mathbb{R}^{n \times n}: \forall x \in \mathcal{D}, V(x) > 0 \wedge \dot{V}(x) \leq 0, 
\end{equation*}
would return a boolean value, i.e. \texttt{True} or \texttt{False}: to obtain the synthesised $V(x)$ function, we remove the existential quantifier.



\subsection{The CEGIS Architecture for Lyapunov Function Synthesis}
\label{subsec:cegis-architecture}


We introduce the CEGIS architecture to find Lyapunov functions.  
To better illustrate the methodology, we start by considering linear models (the non-linear case is further discussed in Section \ref{subsec:nl-systems}). 
As mentioned earlier, two components characterise the CEGIS approach: a learner and a verifier. 
The CEGIS architecture takes the system matrix $A$ and outputs a matrix $P$ as the key component of the function $V(x)$,  
verifying the conditions in Eq. \eqref{eq:lyapunov-conditions}. 
We denote by $\bar{P}_i$, $i = 0, 1, 2, \dots$ the \textit{candidate} matrices yet to be verified, i.e. the outputs of the learner. 
As anticipated earlier, referring to Eq. \eqref{eq:V=x^i Pi x^i}, we set $c=1$ and $\mathcal{D} = \mathbb{R}^n \setminus \{0\}$. 

\subsubsection{Verifier}
\label{subsubsec:verifiers}

The scope of a verifier is twofold: generate a counterexample to the validity of the candidate Lyapunov function, or certify its validity over a domain of interest. 
We implement the verifier in Z3.
 %

The methodology to assert the correctness of a Lyapunov function is as follows.
Assume the learner computes  a candidate Lyapunov function $V(x)$ and passes it to the verifier (in case of a linear function, the learner offers a matrix $\bar{P}_i$).
The goal of the verifier is to assert the validity of formula $F$ from \eqref{eq:synthesis-formula} according to the specification $\psi$ in \eqref{eq:lyapunov-conditions}.  
The check is performed by negating $F$: if there exists  a vector $\bar{x}$ that satisfies $\neg F$, it is a counterexample for $F$; if it does not exist, formula $F$ is valid and the candidate Lyapunov function is an actual Lyapunov function. 
The domain $\mathcal{D}$ is encoded as an additional formula. Assume, as an example, the domain is an hyper-sphere of radius one: $\mathcal{D}$ can be written formally as $ \mathsf{d}$: $\lvert \lvert x \rvert \rvert ^2 \leq 1$. The final formula thus results in $\neg F \wedge \mathsf{d}$.  

A counterexample $\bar{x}$ must satisfy the formula 
$V(\bar{x})\leq0 \vee \dot{V}(\bar{x})>0$. 
Reasoning on either condition, it is easy to show that if there exists a counterexample $\bar{x}$ invalidating a matrix $\bar{P}$,
then there exists an infinite number of counterexamples for this $\bar{P}$. 
Thus, particularly for high-dimensional models the  
generation of meaningful counterexamples is crucial to find a Lyapunov function quickly.  
%

Let us denote $\bar{x}_i$, $i = 1, \dots$, the series of counterexamples provided by the verifier and $\bar{P}_i$ the series of candidate Lyapunov function matrices provided by the learner. 
In this setting, the learner proposes the first default candidate matrix $\bar{P}_0$; the verifier will (possibly) provide a counterexample $\bar{x}_0$; 
the learner 
includes $\bar{x}_0$ in the set of constraints (cf. Section \ref{subsubsec:iterative-learners}) and offers a new candidate $\bar{P}_1$. 

In this work, we let Z3 generate counterexamples without any further goals.
However, 
counterexamples can be generated adding constraints, e.g. 
linear independence or orthogonality.  
Intuitively, more constraints might generate ``better'' candidates by the learner, albeit at an increase in computational cost.  

As intuition suggests, if we were to work with models having a diagonal matrix $A$, 
then the synthesis of diagonal candidates $\bar{P}_i$ and of a diagonal solution ${P}$ would reduce the number of variables needed, thus speeding up the computation. 
As such, if $A$ is not diagonal but diagonalisable, the algorithm pre-computes the system diagonalisation and feeds it to the CEGIS architecture returning a matrix $P$ for the diagonal system, which is then converted to a solution for the original model. 

\smallskip

\subsubsection{Learner}
\label{subsubsec:iterative-learners}

A learner is the CEGIS component designated to suggest a candidate solution for the problem under consideration. 
Within our framework, a learner solves linear inequalities derived
from $F(V(\bar{x}))$
as per Eq. \eqref{eq:synthesis-formula}, while memorising the set of counterexamples $\{ \bar{x}_i  \mid \neg F (\bar{x}_i) \}$ generated by the verifier.  
Whilst the verifier works over continuous domains, 
note that the learner only considers a \textit{finite} number of points to synthesise the candidate Lyapunov function. 
At each iteration $i$, the learner is tasked to solve $2 i$ linear inequalities: $i$ inequalities for $V\geq0$ and $i$ for $\dot{V}\leq 0$ -- this is two inequalities per counterexample, so a set of useful counterexamples is vital to achieve efficiency. 

We implement two learners, for comparison: 1) a numerical and 2) a Z3-based learner. 
However, our CEGIS architecture can in principle accommodate any learner. 
The first learner uses Gurobi \cite{gurobi}, a fast, commercial optimisation solver for, among
others, linear and quadratic programming problems, supporting continuous variables.    
Notice that the synthesis is a linear program: variables $p_{i, j}$, the entries of matrix $P$, appear linearly within the inequalities in $F(V(\bar{x}_i))$. 
Gurobi is thus expected to outperform an SMT solver in this specific task. 
However these variables do not represent real numbers, but floating point numbers that are approximated at machine precision.  
The second learner instead employs Z3, which is numerically sound and not affected by machine precision.  
Z3 solves an SMT instance to synthesise $V(x)$: 
it asserts the satisfiability of Eq. \eqref{eq:synthesis-formula} $F(V(\bar{x}_i))$ for all collected counterexamples $\bar{x}_i$.

As mentioned earlier, the number of inequalities to be solved depends on the number of counterexamples, which can grow to be quite large. 
Whilst the verifier ought to generate useful counterexamples, the learner is optimised to output a matrix $\bar{P}_i$ that is easy to handle. 
%
%
%
The comparison between a numerical learner (running on Gurobi) and a sound one (based on Z3) 
shows that the compromise between speed and soundness results is evident (cf. Section \ref{sec:case-study}). 
Z3 is sound, yet slower when compared to the numerical learner. 

Z3 offers an incremental feature to the learner. 
During each CEGIS loop, on the verification side 
the memory is cleared from the previous constraints as the verifier re-initialises the verification problem with a new candidate $V(x)$.
On the other hand, the learner keeps the previous synthesis instance adding a new constraint related to the latest counterexample.
This incremental approach reduces the computational effort, as the learner does not initialise a new problem for every CEGIS loop. 

\subsection{Lyapunov Function Synthesis for Non-linear Models}
\label{subsec:nl-systems}

The problem of synthesizing Lyapunov functions and their region of validity for a general non-linear system 
$\dot{x} = f(x(t))$ is approached via linearisation or via direct computation.  

The linearisation approach consists of three steps for the learner: 
we first linearise the $f(x(t))$, obtaining 
$$
\dot{\tilde{x}}(t) = A_L \tilde{x}(t),
$$
where $A_L$ is the Jacobian of $f(x(t))$ evaluated at $x_e$;
we then compute matrix $P$ -- and quadratic Lyapunov function $V(x) = x^T P x$ -- on the linearised system; 
finally, we find $\mathcal{R}$, defined as the set in which the linear Lyapunov function is valid. 
%
Next, we detail the synthesis of region $\mathcal{R}$. 
Consider, without loss of generality, an autonomous non-linear system with (at least one) equilibrium point $x_e = 0$.
Assume the CEGIS procedure is successful, i.e. it finds a Lyapunov function $V_L(x) = x^T P x$ that guarantees the asymptotic stability of system $\dot{\tilde{x}} = A_L \tilde{x}$ around $x_e$. 
We now compute the region where $V_L(x)$ guarantees stability with the original system, i.e. $\dot{x} = f(x)$.
In view of the existence of $V_L(x)$ and by definition of linearisation, there exists a neighbourhood of the origin $\mathcal{B}_0$ in which the derivative of the Lyapunov function $\dot{V}(x)$ is non-positive; formally such set is defined as 
\begin{equation*}
\mathcal{B}_0 = \{ x \in \mathbb{R}^n {\setminus \{0\}} \mid \dot{V}(x) \leq 0 \}, 
\end{equation*}
where $\dot{V}(x)$ is computed on the original system, namely
\begin{equation*}
\dot{V}(x) = \nabla V_L(x) \cdot f(x).
\end{equation*}
Let us define 
the boundary of $\mathcal{B}_0$ as 
$
\partial \mathcal{B}_0 = \{ x \in \mathbb{R}^n {\setminus \{ 0\} } \mid \dot{V}(x) = 0 \}.
$
This set may be composed by single points or regions of the state space: in this case, we find $r$, the closest point to the equilibrium that belongs to $\partial \mathcal{B}_0$, as
\begin{equation*}
r = \min_{x \in \partial \mathcal{B}_0} \sum_l x(l)^2.
\end{equation*}
We finally compute region $\mathcal{R}$ as a hyper-sphere of radius $r$, 
\begin{equation}
\mathcal{R} = \{ x \in \mathbb{R}^n {\setminus \{0\} } \mid \| x \|^2 < r^2  \},
\label{eq:stability-region}
\end{equation}
defining the region where the Lyapunov function is valid. 
Finally, region $\mathcal{R}$ is tested with the verifier: formula $F(V(x))$ from Eq. \eqref{eq:synthesis-formula} is passed to Z3 with $\mathcal{D} = \mathcal{R}$. 
Our implementation uses a numerical optimisation technique to compute a value for $r$ that is passed to Z3, as Z3 does not natively handle non-linear optimisation problems. 
With this selection, the region $\mathcal{R}$ represents a sound under-approximation of the maximal stability region.
The linearisation method is used in view of its rapid and effective synthesis capability. 
However, it produces a Lyapunov function that does not ensures global stability when one of the eigenvalues of $A_L$ is equal to zero. 
This is a well-known limitation of the linearisation, which suggests a more formal approach, called \textit{direct computation method}.

The direct computation method, as the name suggests, analytically computes $V(x)$ and $\dot{V}(x)$ from a template $V(x)$ as in Eq. \eqref{eq:V=x^i Pi x^i}.  
The learner is tasked with resolving conditions $\psi$ obtained by a light relaxation of the two inequalities in \eqref{eq:lyapunov-conditions}, namely
\begin{equation*}
V(x) \geq 0,
\quad 
\dot{V}(x) = \nabla V(x) \cdot f(x) \leq 0.
\end{equation*}
Note that the first inequality is not strict: this relaxation allows for a faster computation of a candidate. 
The verifier, on the other hand, produces counterexamples for $V(x) > 0$, thus retaining soundness of the overall procedure.   
The CEGIS framework allows the separation between synthesis and verification. 
So whilst the learner might propose candidates being completely independent from domain $\mathcal{D}$,  
the verifier is responsible to assert or to find the domain of validity $\mathcal{D}$. 
Our implementation establishes that 
at first the verifier checks the validity of $V(x)$ on the whole state space $\mathcal{D} = \mathbb{R}^n$; if the computation is not successful -- namely, the computational time is greater than a predefined timeout -- the verifier checks its validity over a smaller region, e.g. $\mathcal{D} = [-1, 1]^n$, and so on. 
If also this program fails, the algorithm returns an empty $V(x)$. Recall that our algorithm is in general not complete - indeed, 
consider the trivial problem of the synthesis of a Lyapunov function for an unstable system, which is not possible: in this case, the CEGIS procedure will surely return an empty $V(x)$. 


\subsection{Lyapunov Function Synthesis for Parametric Models}
\label{subsec:param-sys-synthesis}

Parametric models represent a challenge for both sound and numerical solvers.
Let us remark that both Gurobi and Z3 cannot synthesise functions in the presence of uncertainty, 
whereas Z3 can provide counterexamples using one or more variables as fixed parameters, using the quantifier \texttt{ForAll}. 


Let us consider variable $x$, a parameter $\mu$ and a formula $\psi(x, \mu)$: 
Z3 can find a counterexample for all values of $\mu$ by validating \texttt{ForAll($\mu$, $\psi$)}. 
If $\mu$ belongs to a range $[l, u]$, Z3 can find a counterexample by checking \texttt{$\psi$ $\wedge$ $\mu\geq l$ $\wedge$ $\mu \leq u$}. 
This provides a counterexample ($\bar{x}, \bar{\mu}$) for $x$ and $\mu$, respectively.

The synthesis procedure is split into two steps, in view of the inability of Z3 and Gurobi to propose parametric solutions. 
The first step synthesises a candidate Lyapunov function solely using the constraint $V(x) > 0$, in which no parameter appears. 
The second step evaluates the constraint $\dot{V} \leq 0$ to propose a parametric Lyapunov function exploiting the results from the first step. 
The following example details the procedure.  
\begin{example}
Consider a two-dimensional linear parametric system  \cite{sankaranarayanan2013lyapunov} and a candidate Lyapunov function 
\begin{equation*}
\begin{cases}
\dot{x} = y \\
\dot{y} = - (2+\mu) x - y
\end{cases}, 
\quad V(x, y) = p_{1} x^2 + p_2 y^2.
\end{equation*}
Assume 
the first guess of the learner is invalid, i.e. the verifier finds a counterexample for the validity of $V(x, y)$.
The counterexample $(\bar{x}, \bar{y})$ is then sent to the learner. 
The synthesis procedure is split into two steps:
the first step entails the synthesis solely accounting for $V(\bar{x}, \bar{y})>0$. The learner is tasked to solve 
$$
V(\bar{x}, \bar{y}) = p_1 \bar{x}^2 + p_2 \bar{y}^2 > 0,
$$
where $p_1$, $p_2$ are the variables of the inequality. 
The learner will propose values $\bar{p}_1$ and $\bar{p}_2$ satisfying the inequality. 
The second step removes one of the synthesised $\bar{p}_i$, e.g. $\bar{p}_1$, in order to re-synthesise it including the parameters found in $\dot{V}$. 
In practical terms, the expression of $\dot{V}$ is evaluated at $\bar{x}$, $\bar{y}$ and $\bar{p}_2$, as
$$
\dot{V} = 2 p_1 \bar{x} \bar{y} - 2 \bar{p}_2 \bar{y}^2 - 2(\mu+2) \bar{x} \bar{y} \leq 0
\Longrightarrow
p_1 \leq \bar{p}_2 \left(  \frac{\bar{y}}{\bar{x}} + 2 + \mu \right).
$$
We choose the value $p_1$ that satisfies the equality. 
The candidate Lyapunov function thus results in $V(x, y) = \bar{p}_2 \left(  \frac{\bar{y}}{\bar{x}} + 2 + \mu \right) \cdot x^2 + \bar{p}_2 \cdot y^2$.
This procedure holds as long as $\bar{x} \neq 0$: if this is not the case, we can either choose to synthesise a new value for $p_2$ or simply maintain the numerical values obtained after the first step. 
In the latter case, once the candidate Lyapunov function is passed to the verifier, a new counterexample will be generated and the procedure can be repeated until a parametric Lyapunov function is found and verified.
Another possible approach is based on the mixed-terms removal: $p_1$ is synthesised so that the terms carrying $\bar{x} \bar{y}$ cancel out. 
Further, the choice of $p_1$ satisfying the equality is arbitrary: we can add a negative constant to its value to solve the strict inequality instead.
Finally, more than one parameter $\bar{p}_i$ can be removed in the second step: this can spread the parametric coefficients among more than one ${p}_i$. 
However, this is likely to increase the computational cost in view of the inequality being a function of more than one variable.
\hfill $\square$
\end{example}


\section{Case Studies and Experiments}
\label{sec:case-study}


In this Section we outline a few experiments to challenge the validity of our approach. 
Our technique is coded in Python 2.7 \cite{python-web}, 
using external libraries as the numerical solver Gurobi and the SMT solver Z3 (cf. Section \ref{sec:formal-methds}). 
Specifically, we compare two CEGIS architectures: 
\begin{enumerate}
	\item Gurobi learner and Z3 verifier,
	\item Z3 learner and Z3 verifier, 
\end{enumerate}
later denoted as \textit{Gurobi-CEGIS} and \textit{Z3-CEGIS}, respectively, 
against the optimisation toolbox SOSTOOLS.  
Whilst Z3 is an efficient verifier,  
it carries the weight of exact representations. 
We therefore compare its use within the learner to that of a numerical solver such as Gurobi - recall that 
the learner does not need to be sound.  
A relevant feature of the synthesis procedure is its \textit{linearity} in the entries of matrix $P$: 
we expect an efficient LP solver to outperform an SMT solver. 
As such,  
we study the expected tradeoff between speed and precision. 
As specified earlier, the initial candidate for the learner $\bar{P}_0$ is arbitrary: 
we challenge the procedure by setting $\bar{P}_0= - I$, which does not satisfy the first positivity condition for Lyapunov functions,  
thus showing that even with an ill-suited initial guess the procedure can rapidly synthesise a valid Lyapunov function.  
SOSTOOLS is a sum-of-squares optimisation toolbox available for MATLAB, equipped with the solver SeDuMi \cite{sedumi}.
It can be used to solve a wide range of problems, from mixed continuous-discrete optimisations to finding Lyapunov functions for polynomial dynamical systems.  

We consider linear, non-linear and parametric ODEs with the origin as (one of) the equilibrium(a), 
and aim to obtain a Lyapunov function guaranteeing the stability of such equilibrium point. 
The procedure entails the following steps: 
\begin{enumerate}
    \item[a)] a function $f(x)$, $x \in \mathbb{R}^n$, is fed as the input;
	\item[b)] a Lyapunov function $V(x)$, as in Eq. \eqref{eq:V=x^i Pi x^i}, is computed;
    \item[c)] in the linearisation case, the stability region $\mathcal{R}$ in Eq. \eqref{eq:stability-region} for $V(x)$ is found. 
\end{enumerate}  
Let us emphasise that Z3 is unable to fully handle non-polynomial terms, which represents the only limitation of our approach.  
Unlike most of the literature, counterexamples are not limited to a finite set but searched over the whole $\mathbb{R}^n$. 


Linear models are certainly an easier task than polynomial systems. 
The study with linear models focuses mainly on the scalability of the method, encompassed by the average and maximum/minimum computational time, and the number of iterations performed. 
We generate $N = 100$ random linear models of dimension $n \in [3, 10]$.
For each linear system, 
the entries of matrix $A$ range within $[-1000, 1000] \in \mathbb{R}$. 
For each test we set $c=1$ (cf. Eq. \eqref{eq:V=x^i Pi x^i}), namely we impose a quadratic structure to the Lyapunov function, 
and collect the number of iterations of the procedure, i.e. the number of counterexamples needed to compute a valid Lyapunov function, 
and the total elapsed time. 
Recall that the initial synthesiser's candidate is $\bar{P}_0 = -I$, which challenges the reliability of our method with a bad initial condition. 
A 180 seconds timeout is set for every run. 
Results comparing the numerical learner using Gurobi and the sound learner using Z3 are reported in Table \ref{tab:results-compared-linear}. 
The average values, as well as the minimum and maximum value among the $N$ random systems, are computed on the synthesis tests that have not timed out. 
The number of timed out procedures are also listed in the Table.


With regards to non-linear and parametric models, 
we assess our approach over a suite of examples taken from related work on Lyapunov function synthesis 
\cite{papachristodoulou2002construction}, 
\cite{prajna2004sostools}, \cite{papachristodoulou2018sostools},
\cite{sankaranarayanan2013lyapunov}, 
which are reported in the following. 
The value $c$ from Eq. \eqref{eq:V=x^i Pi x^i} is set heuristically as \texttt{ceil}($d$/2), where $d$ is the order of the system   
(this choice follows the common interpretation of Lyapunov maps as storage functions).   
Due to ease of implementation, only Z3-CEGIS performs the synthesis with $c>1$ and in the case of parametric models. 
Results in terms of computational time and iterations are reported in Table \ref{tab:nl-results}.  
%
Experiments are run on a 4-core Dell laptop with Fedora 30 and 8GB RAM.

\smallskip



\smallskip


\begin{example} 
\label{ex:nl-6dim}
Consider the model \cite{papachristodoulou2002construction}
\begin{equation*}
\begin{array}{rclcrcl}
\dot{x}_1 & = & -x^2_1-4x^3_2-6x_3x_4 ,
& \quad &
\dot{x}_4 & = &  x_1x_3+x_3x_6-x^3_4, \\ 
\dot{x}_2 & = & -x_1-x_2+x^3_5,
& \ &
\dot{x}_5 & = &  -2x^3_2-x_5+x_6,
\\
\dot{x}_3 & = & x_1x_4-x_3+x_4x_6,
& \ &
\dot{x}_6 & = & -3x_3x_4-x^3_5-x_6.
\end{array}
\end{equation*}
Z3-CEGIS finds the Lyapunov function  
$
V(x) =  2 x_1^2 + 4 x_2^4 +  x_3^2 + 11x_4^2 + 2 x_5^4 + 4 x_6^2, 
$ 
ensuring stability over the whole state space. 
SOSTOOLS fails to find a $2^{nd}-$ or $4^{th}-$order Lyapunov function for this model.  \hfill $\square$
\end{example}

\smallskip

\begin{example} 
\label{ex:nl-2dim-simple}
Consider the model \cite{sankaranarayanan2013lyapunov} 
\begin{equation*}
\begin{cases}
\dot{x} = - x^3 + y &  \\
 \dot{y} = -x -y. & 
\end{cases} 
\end{equation*}
Gurobi-CEGIS finds the Lyapunov function 
$
V(x) = 
5 \cdot 10^{-5} x^2 + 5 \cdot 10^{-5} y^2,
$
 whereas Z3-CEGIS finds $V(x) = 0.5 x^2 + 0.5 y^2$, 
both ensuring global  stability. 
The linearised Gurobi-CEGIS finds $V(x) = 3.2 \cdot 10^{-3} x^2 + 3.2 \cdot 10^{-3} y^2$,  
whereas SOSTOOLS finds $V(x) = 0.7844(x^2+y^2)$, also ensuring stability over the whole state space. 
\hfill $\square$
\end{example}

\smallskip

\begin{example} 
\label{ex:nl-3dim-frac}
Consider the system \cite{papachristodoulou2018sostools} 
\begin{equation*}
\begin{cases}
\dot{x}_1 = - x_1^3 - x_1 x_3^2, \\
\dot{x}_2 = -x_2 - x_1^2 x_2, 
\\
\dot{x}_3 = -x_3 - \dfrac{3 x_3}{x_3^2 + 1} + 3 x_1^2 x_3.
\end{cases} 
\end{equation*}
Note that the term $x_3^2+1$ is always non-negative, therefore we can consider $\dot{V}(x) \cdot (x_3^2+1)\leq 0$.
Gurobi-CEGIS finds the Lyapunov function 
$
V(x) = 32 \cdot 10^{-4} x_1^2 + 32 \cdot 10^{-4} x_2^2 + 8 \cdot 10^{-4} x_3^2,
$
whereas 
Z3-CEGIS finds $V(x) = 
3 x_1^2 +  x_2^2 +  x_3^2$,  
and finally SOSTOOLS finds the function $V(x) = 6.659x1^2 + 4.628x2^2 + 2.073x3^2$, 
all ensuring global stability. 
\hfill $\square$
\end{example}

\smallskip

\begin{example} 
\label{ex:nl-2dim-hard}
Consider the system \cite{sankaranarayanan2013lyapunov} 
\begin{equation*}
\begin{cases}
\dot{x}= - x - 1.5 x^2 y^3,
\\
\dot{y}=-y^3 + 0.5x^3 y^2.
\end{cases}
\end{equation*}
Z3-CEGIS finds $V(x) = 1/3 x^2 + y^2$, valid on the whole $\mathbb{R}^2$, 
whereas SOSTOOLS finds $V(x) = 0.4707 x^2 + 1.412 y^2$, with a stability region of radius $r=68$. 
Gurobi-CEGIS returns an error, as it finds
$V(x) = 1.00066454641347 x^2 + 2.99933545358653 y^2$ that is \textit{not} a valid Lyapunov function.
The correct solution, $V(x) = x^2 + 3y^2$, can not be attained in view of lack of convergence of the optimisation algorithm.
On the other hand, the linearised Gurobi-CEGIS delivers $V(x) = 32 \cdot 10^{-4} x^2 + 2 \cdot 10^{-4} y^2$ with a radius $r = 1.25$.
\hfill $\square$
\end{example}

\smallskip

\begin{example} 
\label{ex:nl-4dim}
Consider the system \cite{sankaranarayanan2013lyapunov}:
\begin{equation*}
\begin{array}{rclcrcl}
\dot{x}_1 & = & -x_1+x^3_2-3x_3x_4,
& \quad & 
\dot{x}_3 & = & x_1 x_4 - x_3, 
\\
\dot{x}_2 &  = &  - x_1 -x^3_2,
& \quad & 
\dot{x}_4 & = & x_1 x_3 - x^3_4.
\end{array}
\end{equation*}
Z3-CEGIS finds the Lyapunov function $V(x) = 2 x_1^2 + x_2^4 + 3201/1024 x_3^2 + 2943/1024 x_4^2 $, ensuring global stability. 
SOSTOOLS, on the other hand, finds a complex $4^{th}$ order polynomial, omitted here for brevity, with a stability region that is hard to characterise analytically. 
\hfill $\square$
\end{example}

\begin{example}
\label{ex:param}
Consider the parametric linear system \cite{sankaranarayanan2013lyapunov} 
\begin{equation*}
\begin{cases}
\dot{x} = y, 
\\  
\dot{y} = -(2+\mu) x - y,
\end{cases}
\end{equation*}
where $\mu \in (-2, 5]$.
Z3-CEGIS discovers the Lyapunov function $V(x) = (\mu+2)x^2 + y^2$, 
ensuring stability on the whole state space.  
On the other hand,
SOSTOOLS fails to find a solution when setting $V(x, \mu)$ to be independent from, linear in, or quadratic in $\mu$. 
\hfill $\square$
\end{example}

\begin{example}
\label{ex:param-large}
Consider the parametric system \cite{sankaranarayanan2013lyapunov} 
\begin{equation*}
\begin{cases}
\dot{x} = -(1+\mu_1)x+ (4+\mu_2)y, 
\\
\dot{y} = -(1+\mu_3) x - \mu_4 y^3,
\end{cases}
\end{equation*}
where $\mu_i \in [0, 100]$ for $i = 1, \dots 4$.
Z3-CEGIS discovers the Lyapunov function $V(x) = \dfrac{\mu_3+1}{\mu_2+4} x^2 + y^2$ that asserts stability on the whole state space, 
whereas SOSTOOLS can not find a solution considering $V(x)$ independent from, linear in, or quadratic in $\mu_i$, where $i=1, \dots, 4$. 
\hfill $\square$
\end{example}

As expected, Gurobi is faster than Z3 in terms of iterations and computational time. The gap becomes larger with a high-dimensional system, as the SMT learner does not implement any optimisation techniques.
The  Z3-CEGIS synthesis is performed via an SMT call, 
which grows in complexity as the number of constraints -- related to the number of counterexamples -- increases. 
Gurobi, on the other hand, using optimisation techniques converges faster to a candidate solution that is closer to the actual solution. 
Our approach outperforms SOSTOOLS in terms of computational time, and it is able to handle parametric and complex models.  

Notice that the coefficients of the Lyapunov function synthesised by Gurobi are small in magnitude, as the linear programming problem can encompass the minimisation of coefficients in its setup.  
On the other hand those obtained from Z3 (rational fractions) are arguably more interpretable. 
A very interesting result comes from Example \ref{ex:nl-2dim-hard}. 
Gurobi-CEGIS converges towards the correct Lyapunov function, yet it can not reach the exact numerical values in view of the algorithmic precision. 
Gurobi numerical guidelines \cite{gurobi} suggest that, as a rule of thumb, the ratio of the largest to the smallest coefficient of the LP problem should be less than $10^9$. In our setting, the coefficients are the 
counterexamples found by Z3, which might require higher precision. In this case, the issue is (probably) caused by a counterexample $\bar{x} \simeq [-755145, 1/8]$, where the first element is actually represented as a (very long) ratio between two integers. The ratio between the two $\bar{x}$ coefficient is in the order of $10^7$. 
Roughly speaking, the counterexamples generated by Z3 depend on the complexity of the tested model: a high-order system might generate numerically ill-conditioned  counterexamples, as this example shows.
It is also significant how the numerical algorithm tries to converge to a correct solution. 
The first candidate Lyapunov function results in $V(x) = 1.07079661938449 x^2 + 2.92920338061551 y^2$ and it takes 99 counterexamples to reach the final value (cf. Example \ref{ex:nl-2dim-hard}), until the procedure stops, resulting in an infeasible problem. 
Even enveloping the numerical values with the Python Sympy objects \texttt{Rational}, \texttt{Decimal}, \texttt{Fraction}, or the function \texttt{simplify} do not help in this context, the limitation being Gurobi's numerical precision.

\begin{table*}
\centering
\begin{tabular}{c|c|c}
$n$ & Gurobi-CEGIS & Z3-CEGIS \\ \hline
  \begin{tabular}{c}
     \\
   3 \\
   4 \\
   5 \\
   6 \\
   7 \\
   8 \\
   9 \\
   10 \\
  \end{tabular}
  &
  \begin{tabular}{c|c|c}
  Iterations & Time [sec] & Oot  \\ \hline
  3 [3, 3] & 0.48 [0.33, 0.77] & -- \\
  3.10 [3, 4] & 0.53 [0.36, 1.20] & --  \\
  4.15 [4, 5] & 1.33 [1.08, 1.97] & -- \\
  6.99 [4, 10] & 3.88 [2.41, 4.97] & -- \\
  8.56 [4, 12]  & 12.64 [2.9, 62.3] &  -- \\
  9.14 [3, 13] & 21.50 [3.9, 114.16] & 1 \\
  15.72 [3, 32] & 29.98 [3.87, 78.5] & 2 \\
  18.45 [3,41] & 40.63 [6.17, 46.65] & 5
  \end{tabular}
  & 
  \begin{tabular}{c|c|c}
  Iterations & Time [sec] & Oot \\ \hline
  3.03 [3, 4] & 0.49 [0.4, 0.70] & --\\
  5.93 [4, 7] & 0.68 [0.54,1.07] & -- \\
  7.38 [5, 12] & 1.67 [1.10, 3.03] & --\\
  9.10 [6, 10] & 7.48 [2.40, 54.44] & --\\
  12.88 [5, 17] & 17.63 [5.41, 20.3] & 1\\
  16.2 [3, 25]  & 23.91 [4.05, 35.08]  & 1 \\
  22.47 [4, 35] & 34.41 [5.67, 48.96]  & 5 \\
 27.25  [5, 47] & 44.63 [6.32, 101.2]  & 7 \\     
  \end{tabular} \\ 
\end{tabular} 
\\
\caption{Comparison between Gurobi-CEGIS and Z3-CEGIS over $n$-dimensional linear models. 
The first values are the average performance on the $N=100$ randomly generated models, and within brackets the minimum and maximum values. Oot is the number of runs (out of $N$) not finishing after $180$ [sec].}
\label{tab:results-compared-linear}
\end{table*}





\begin{table}
\centering
\begin{tabular}{c|c|c|c}
Example \# & Gurobi-CEGIS & Z3-CEGIS & SOSTOOLS \\ \hline
\begin{tabular}{c}
 \\
\ref{ex:nl-6dim} \\
\ref{ex:nl-2dim-simple}   \\
\ref{ex:nl-3dim-frac}  \\
\ref{ex:nl-2dim-hard}  \\
\ref{ex:nl-4dim} \\
\ref{ex:param}  \\
\ref{ex:param-large}
\end{tabular}
	& 
\begin{tabular}{c|c}
Time [sec] & Iterations \\
\hline
 -- & --  \\
 0.32 & 2  \\
 0.37 & 4  \\
 0.16 & 2  \\
 -- & --  \\
 -- & --  \\
 -- & --
\end{tabular}
	&
\begin{tabular}{c|c}
Time [sec] & Iterations \\
\hline
 18.38 & 4 \\
 1.27 & 5  \\
 0.60 & 3 \\
 0.27 & 2 \\
9.26 & 3 \\
 0.14 & 3 \\
 0.23 & 3
\end{tabular}
	&
\begin{tabular}{c}
Time [sec]  \\
\hline
 --  \\
 3.66   \\
 4.38  \\
 3.83 \\
21.31  \\
 --  \\
 -- 
\end{tabular}
\end{tabular}
\caption{ 
Comparison between Gurobi-CEGIS, Z3-CEGIS and SOSTOOLS for non-linear models (see Examples description in main text). The result for Gurobi-CEGIS in Example \ref{ex:nl-2dim-hard} is obtained via linearisation. }
\label{tab:nl-results}
\end{table}


\section{Conclusions and Future Work}
\label{sec:concl}

In this work, we have studied the problem of automated and sound synthesis of Lyapunov functions. 
We have exploited a CEGIS framework, equipped with a sound verifier (the Z3 SMT solver) and with either a numerical LP solver (Gurobi) or a sound (Z3) learner. 

We have provided a simple -- yet effective --  methodology to synthesise Lyapunov functions for linear, polynomial and parametric systems and shown
evidence of scalability and reliability of our method using benchmarks from the literature.   
We have in particular synthesised quadratic Lyapunov functions for linear models and verified their validity on the whole state space. 
We have tackled non-linear models 
following two approaches: either 
1) the computation of Lyapunov functions over the linearised system and the synthesis of its validity region; or 
2) the direct computation of a higher-order Lyapunov function. 

Future work includes the implementation of synthesis techniques for Gurobi-CEGIS for high-order and parametric models, 
together with the study of optimisation techniques for the synthesis in Z3-CEGIS: 
the tuning of the SMT solvers leaves much room, for example in order to provide insightful counterexamples or to additionally optimise an objective function. 
Further, we aim at embedding CEGIS with neural networks (as function approximators) to replace the learner, 
whilst maintaining the verification in the hands of an SMT solver - this approach has been recently pursued also in \cite{ChangRG19}. 



\bibliographystyle{IEEEtran}
\bibliography{main.bib}


\newpage

{\small\medskip\noindent{\bf Open Access} This chapter is licensed under the terms of the Creative Commons\break Attribution 4.0 International License (\url{http://creativecommons.org/licenses/by/4.0/}), which permits use, sharing, adaptation, distribution and reproduction in any medium or format, as long as you give appropriate credit to the original author(s) and the source, provide a link to the Creative Commons license and indicate if changes were made.}

{\small \spaceskip .28em plus .1em minus .1em The images or other third party material in this chapter are included in the chapter's Creative Commons license, unless indicated otherwise in a credit line to the material.~If material is not included in the chapter's Creative Commons license and your intended\break use is not permitted by statutory regulation or exceeds the permitted use, you will need to obtain permission directly from the copyright holder.}

\medskip\noindent\includegraphics{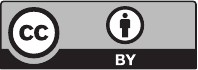}

\end{document}